\documentclass[10pt]{article}
\usepackage{amssymb,amsfonts,amsmath,latexsym}
\newcommand{\be}{\begin{equation}}
\newcommand{\ee}{\end{equation}}
\newcommand{\bea}{\begin{eqnarray}}
\newcommand{\eea}{\end{eqnarray}}
\newcommand{\ba}{\begin{array}}
\newcommand{\ea}{\end{array}}
\newcommand{\bt}{\begin{tabular}}
\newcommand{\et}{\end{tabular}}

\newcommand{\sump}{\mathop{{\sum}'}}
\newcommand{\prodp}{\mathop{{\prod}'}}
\newcommand{\ti}{\tilde}
\newcommand{\wti}{\widetilde}
\newcommand{\fr}{\frac}
\newcommand{\ci}{\cite}
\newcommand{\cl}{\centerline}
\newcommand{\bs}{\bigskip}

\newcommand{\vs}{\vspace}

\newcommand{\en}{\eqno}

\newcommand{\bbib}{\begin{thebibliography}}
\newcommand{\ebib}{\end{thebibliography}}

\newcommand{\lra}{\leftrightarrow}

\newcommand{\ol}{\overline}
\begin{document}

\cl{\bf EFFECTIVE CONDUCTIVITY}
\cl{\bf OF SELF-DUAL RANDOM HETEROPHASE SYSTEMS}

\bs

\cl{\bf S.A.Bulgadaev \footnote{e-mail: bulgad@itp.ac.ru}}

\bs
\cl{Landau Institute for Theoretical Physics}
\cl{Chernogolovka, Moscow Region, Russia, 142432}

\bs

\begin{quote}
\footnotesize{
The duality and other symmetry properties  of the effective conductivity $\sigma_{e}$
of the classical two-dimensional isotropic randomly inhomogeneous heterophase
systems at arbitrary number of phases $N$ are discussed. A new approach for a obtaining
$\sigma_{e}$  based on a duality
relation is proposed. The exact values of $\sigma_{e}$
at some special sets of the partial parameters are found.
The  explicit basic solutions of the duality relation, connected with the
higher moments and satisfying all necessary requirements, are found at arbitrary
values of partial parameters. It is shown that one of them can describe $\sigma_{e}$
for systems with a finite maximal characteristic scale of the
inhomogeneities in a wide range of parameters. The other solution, connected with
a mean conductivity describes $\sigma_{e}$ of the introduced random parquet model
of $N$-phase randomly inhomogeneous medium in some mean field like approximation.
The comparison with the known effective medium approximation and crossover to
the continuous smoothly inhomogeneous case are also discussed.
}
\end{quote}

\bs

\bs
\cl{\bf 1. Introduction}
\bs

The properties of the electrical transport of
classical macroscopically inhomogeneous (randomly or regularly) heterophase systems,
consisting of $N \;(N \ge 2)$  phases with different
conductivities  $\sigma_i \; (i =1,2,...,N),$ always have had  a great significance
for practice (see, for example \ci{1,2}). Their importance revived in the last decades
under investigation of various new materials, but now the inhomogeneities have
more smaller scales (from microscopic and mesoscopic till macroscopic)\ci{3,4}.
The main problem in a theory of the electrical transport of heterophase systems is a
calculation of the effective conductivity $\sigma_{e}$ at arbitrary
partial conductivities and phase concentrations. Though some general results have been
obtained for weakly inhomogeneous media \ci{1}, this problem remains
unsolved for inhomogeneous media with strong inhomogeneous fluctuations,
which can be connected with strong irregularities
as well as with large differences of partial conductivities, due to a difficulty
of the averaging over such fluctuations.
In order to obtain any explicit expression for $\sigma_{e}$  a mean field like
procedure, the effective medium (EM) approximation, has been offered
many years ago \ci{5,6}. It is applicable for all dimensions and
describes well $\sigma_{e}$ of heterophase systems for weak and
moderate inhomogeneities for systems, admitting the random
resistor network (RRN) representation \ci{7,8}.
Unfortunately, the RRN model, being a discretised version of the continuous
medium, is not a good representation for all continuous media with moderate
inhomogeneities \ci{9}.
Moreover, already in 3-phase case, the EM approximation gives for
$\sigma_{e}$ a complicated algebraic expression, which is
practically untractable (see below section 4 and Appendix C).
For this reason one needs other methods and approximations, which will be more
adequate for continuous random inhomogeneous media.
Fortunately, in two dimensions there is another possibility to treat the problem.
As is known, the two-dimensional systems have an exact
dual symmetry, which is a consequence of a duality between potential and
divergenceless fields in two dimensions.
In case of two-phase  isotropic self-reciprocal systems it gives the exact relation,
connecting  $\sigma_{e}$ at the inversed values of the partial conductivities
$\sigma_i,$ \ci{10,11}
$$
\sigma_{e}(\sigma_1,\sigma_2,x)
\sigma_{e}(\sigma_0^2/\sigma_1,\sigma_0^2/\sigma_2,x) =
\sigma_0^2,
\en(1)
$$
where $\sigma_0$ is some arbitrary parameter with a dimensionality of a conductivity,
$x$ is a concentration of one (for example, the first one) phase.
For self-dual systems (i.e. systems, satisfying (1) and symmetrical under the
permutation of the partial parameters $(\sigma_1, x_1) \lra (\sigma_2, x_2)$)
the relation (1) can be represented also in the form,
connecting $\sigma_{e}$ at adjoint concentrations $x$ and $1-x$
$$
\sigma_{e}(\sigma_1,\sigma_2,x) \sigma_{e}(\sigma_1,\sigma_2,1-x) =
\sigma_1 \sigma_2.
\en(2)
$$
This duality relation (DR) allows to find the exact universal,
remarkably simple, formula for $\sigma_{e}$ at equal phase concentrations
$x_1 = x_2 = x_c = 1/2$
$$
\sigma_{e} = \sqrt{\sigma_1 \sigma_2}.
\en(3)
$$

Recently it was shown  that the DR with some additional assumptions can be
used as a basic relation for a determination of $\sigma_e$ of inhomogeneous two-phase
systems, and two different explicit approximate expressions for $\sigma_e$
have been derived \ci{9,12}. They differ from the EM approximation and satisfy all
necessary requirements.
One of these expressions, generalizing the exact
Keller -- Dykhne formula (3) on arbitrary concentrations
$$
\sigma_{e}(\sigma_1,\sigma_2,x) = \sigma_1^x \sigma_2^{1-x},
\en(4)
$$
describes the effective conductivity of self-dual systems with compact inclusions
of one phase into another and is applicable almost everywhere, except the regions,
where one of the partial conductivities $\sigma_i \to 0$ \ci{9,12}. The other expression
for $\sigma_e$  describes $\sigma_e$ of two-phase random parquet model with layered
(or striped) structure. The existence of various expressions for $\sigma_e$ means also
that it is a nonuniversal function, depending on the structure of the inhomogeneities.

In this paper we will study general properties of $\sigma_{e}$ of two-dimensional
self-dual heterophase systems with arbitrary number $N$ of phases and show that the
analogous expressions can be obtained for systems with $N \ge 2$.
Firstly, we will discuss general properties of these systems, including a
duality relation (a section 2), and find the
fixed points of the duality transformation and the corresponding  exact values for
$\sigma_e$ (a section 3).
Then, in the section 4, we consider the EM approximation for $N$-phase systems
and show how the found exact values can be reproduced in this approximation.
In the section 5 we propose
a general solution of the duality (or inversion) relation and obtain some
explicit expressions for $\sigma_e,$ two of which generalize the abovementioned
solutions of $N=2$ case.
Then, two physical models of continuous random inhomogeneous media, whose effective
conductivities coincide with these two expressions, will be presented in the
sections 6 and 7. The first model represents the $N$-phase systems with
compact inclusions of different phases.  For this model
we deduce in the framework of the finite maximal scale averaging
approximation (FMSA approximation) introduced in previous papers \ci{9,12}
the functional self-consistency
equation for  the effective conductivity and find a solution of this equation,
its physical meaning and properties.
The second model, generalizing the two-phase random parquet model
on $N$-phase case, gives in a mean field like
approximation another expression for effective conductivity, coinciding
with the second solution of the duality relation.
These results demonstrate that  the effective conductivity of self-dual heterophase
systems is a nonuniversal function of partial parameters of phases
and depends on the structure of their inhomogeneities.

\bs

\cl{\bf 2. Symmetry and duality properties}

\bs

Since the inhomogeneous heterophase systems can be very different,
we begin with a discussion of some general properties
of the effective conductivity $\sigma_{e}$ of two-dimensional $N$-phase
systems with the partial conductivities $\sigma_i\; (i = 1,2,...,N)$ and
concentrations $x_i\;(\sum_1^N x_i =1).$ For a brevity,
we will denote a set of all partial parameteres, for example, $x_i, \; (i=1,...,N)$
as $\{x\}.$
It was already noted above that there are
continuous heterophase systems and their discretized RRN type models.
Some properties of these systems are different due to a discretization
procedure, which is approximate and rather ambiguous. For this reason
RRN type models are not always a good approximation for continuous media.

The effective conductivity of $N$-phase systems must be a homogeneous (a degree 1)
function of partial conductivities $\sigma_i$ and
satisfies the following natural boundary conditions:

1) $\sigma_e$ of $N$-phase system with some equal partial conductivities
must reduce to $\sigma_e$ of system with smaller number of phases and
the concentrations of the phases with equal conductivities must add;

2) it has not depend on partial $\sigma_i$ and must reduce to the effective
conductivity of $(N-1)$-phase system, if the concentration $x_i = 0$;

3) it must reduce to partial $\sigma_i$, if $x_i = 1.$

As it follows from the general principles $\sigma_e$ must satisfies also the next
inequalities (see, for example, \ci{14})
$$
\langle \sigma^{-1}\rangle^{-1} \le \sigma_e \le \langle \sigma \rangle,
\en(5)
$$
where $\langle ... \rangle$ denotes the averaging with the distribution function
$$
P(\sigma) = \sum_1^N x_i \delta(\sigma - \sigma_i).
\en(6)
$$
In two-dimensional heterophase systems the effective conductivity $\sigma_e$
satisfies also  at arbitrary values of phase concentrations the reciprocity relation,
which generalizes the reciprocity relation of two-phase systems \ci{8,15,16}
$$
\sigma_{e}(\{\sigma\},\{x\})
\tilde \sigma_{e}(\{\sigma_0^2/\sigma\},\{x\}) =
{\sigma_0}^2,
\en(7)
$$
where $\sigma_0$ is an arbitrary constant conductivity, characterizing
a reciprocity transformation, and $\tilde \sigma_{e}$ is the effective conductivity of
the reciprocal system, connected with the initial one.  This relation is a consequence
of a local duality between potential  and divergenceless vector fields
in two dimensions. The value ${\sigma_0}^2/\sigma_i$ can be named as
inversed to $\sigma_i$. A structure of the reciprocal system depends on the type of
initial system. For continuous heterophase media it coincides with the initial one turned on
$\pi/2,$ and for isotropic systems coincides with the initial one \ci{9}.
For RRN type models the reciprocal system is the RRN model on the reciprocal lattice,
which, in general, does not coincide with the initial lattice even for isotropic models.
For example, in two dimensions there are three isotropic regular lattices:
self-reciprocal square  and mutually reciprocal
triangle and honeycomb. For self-reciprocal systems the relation (7) takes the form
$$
\sigma_{e}(\{\sigma\},\{x\})
\sigma_{e}(\{\sigma_0^2/\sigma\},\{x\}) =
{\sigma_0}^2,
\en(8)
$$
which puts some constraints on a possible functional form of $\sigma_{e}.$
The relation (8) can  be written also in the other form
$$
\sigma_{e}(\{\sigma_0^2/\sigma\},\{x\}) =
{\sigma_0}^2/
\sigma_{e}(\{\sigma\},\{x\}),
\en(8a)
$$
which means that $\sigma_{e}$ at the inversed values of the partial
conductivities $\sigma_i$ is equal to the inversed $\sigma_{e}$!
For this reason (8) can be named also the inversion relation (IR) for
the effective conductivity.
The IR has a simple physical sense, connected with two possibilities of
a writing the Ohm law in terms of conductivities $\sigma$ and in
terms of resistivities $\rho = 1/\sigma$:
an effective conductivity of a self-reciprocal system as a function of  partial
conductivities
must coincide with an inversed resistivity as a function of corresponding
partial resistivities (i.e. inversed partial conductivities) $\rho_i =1/\sigma_i$.
Thus, the DR (and IR) connects $\sigma_{e}$ at inversed partial conductivities,
but at the same partial concentrations. An existence of the duality relation (8)
is very important, because it reflects an existence of some hidden symmetries,
which, in principle,  can help to solve the problem (see, for example, \ci{17}).

Another very important property of inhomogeneous heterophase systems is
connected with their geometry and structure. For random systems with the
symmetrical distribution function from (6) and for other heterophase systems,
having all phases with the same  properties,
$\sigma_{e}$ must be
a symmetrical function of a set of pairs of partial arguments $(\sigma_i,x_i)$.
For example, it fulfils for various symmetrical regular heterophase systems and for all
RRN models, but not for systems with unequal fixed forms of different phases .
As a symmetrical function $\sigma_{e}$ must be invariant under all permutations
$$
\sigma_{e}(\{\sigma\},\{x\}) =
\sigma_{e}(P_{ij}(\{\sigma\},\{x\})),
\en(9)
$$
where $P_{ij}$ is the permutation of the i-th and j-th pairs of arguments
$(\sigma_i,x_i) \lra (\sigma_j,x_j)$.
We will call the self-reciprocal symmetrical inhomogeneous systems the self-dual (SD)
ones. Their effective conductivity must satisfy (8) (or (8a)) and (9).
Below we will consider mainly the SD systems.
By tradition and for a clarity we will conserve the name
"a duality relation" for (8). In the general form (8) it takes place for all
two-dimensional continuous inhomogeneous media as regular as random, and for square
lattice RRN model. The DR for $N \ge 3$ in general cannot be represented in a form
connecting $\sigma_{e}$ at adjoint concentrations, because an inversion transformation
can be reduced to the permutation transformation only for 2-phases systems.

An existence of the duality relation (8)  contains a lot of information about $\sigma_{e}.$
One can obtain from it all known general results for $\sigma_{e}.$
For example, differentiating (8) on the parameter $\sigma_0,$ one can show that
$\sigma_e$ satisfies the equation
$$
\sum_1^N \sigma_i \partial_{\sigma_i} \sigma_{e}(\{\sigma\},\{x\})=
\sigma_{e}(\{\sigma\},\{x\}).
\en(10)
$$
It means that $\sigma_e$ is a homogeneous function of a degree 1 of partial
conductivities $\sigma_i.$
For this reason, on the one hand, the DR for $\sigma_e$
can be written as
$$
\sigma_{e}(\{\sigma\},\{x\})\sigma_{e}(\{1/\sigma\},\{x\}) = 1,
\en(8b)
$$
from which it follows  that effectively $\sigma_e$  is a
dimensionless quantity in two dimensions.
On the other hand, it can be represented also in the form
$$
\sigma_{e}(\{\sigma\},\{x\}) =
\sigma_s f(\{z\},\{x\}),
\en(11)
$$
where $\sigma_s$ is some normalizing factor, its choice is determined by
a convenience of consideration, and $z_i =\sigma_i/\sigma_s$.
For our purposes it will be convenient to choose
it in the form $\sigma_s = (s_N)^{1/N},$ where $s_N = \prod_{i=1}^N \sigma_i$
is the N-th elementary symmetric function. Then
$z_i =\sigma_i/(s_N)^{1/N}, \; \prod_{i=1}^N z_i =1$
and a function $f$ is a symmetrical function of $N$ pairs $(x_i, z_i)$
(only $N-1$ of them are independent). This choice
conserves a symmetry of the dimensionless function $f$ and works well for all
$\sigma_i\ne 0.$ Since the inversion transformation
$I_{\sigma_0}$ acts on $z_i$ as (see Appendix A6) $I_{\sigma_0} z_i = 1/z_i,$
the DR for the function $f$ in terms of variables $z$ has the following form
$$
 f(\{z\},\{x\}) f(\{1/z\},\{x\})=1.
\en(12)
$$
The effective conductivity of self-dual $N$-phase systems, when all phases have
the carriers with the same sign of the charge (we will confine ourselves in this paper
by this case), can be represented in a more symmetrical
form. Introducing the new variables $b_i = \ln z_i,\; \sum_1^N b_i = 1$ one obtains
that $\sigma_e$ of such systems
is a pairwise symmetrical function of two similar sets of parameters $\{b\}$ and $\{x\}.$

One can also obtain from (8b) in a weak inhomogeneous case the general Landau-Lifshitz
formula \ci{1}. Representing  the partial conductivities in this limit
in the form
$\sigma_i = \bar \sigma (1+\delta \sigma_i)\; (\sum_1^N x_i \delta \sigma_i =0),$
where $\bar \sigma = \langle \sigma \rangle$ is a mean value
of $\sigma_{e},$ and substituting them into (8b),
one obtains the  Landau-Lifshitz
formula \ci{1}, which has in a case of $N$-phase systems the form
$$
\sigma_{e}(\{\sigma\},\{x\}) =
\bar \sigma \left(1 - \fr{1}{2}
\sum_{i=1}^N x_i\delta \sigma_i^2 \right).
\en(13)
$$
This means that the formula (13) is a consequence of general properties of self-dual
heterophase systems.

\bs

\cl{\bf 3. Fixed points and exact values for self-dual  systems}

\bs

There is another possibility to use the DR (or IR) of inhomogeneous self-dual systems.
Giving to $\sigma_0$ different
values, one can obtain various relations, connecting $\sigma_{e}$ at
different values of its arguments. The DR (8) can be used also for obtaining
the exact equations for $\sigma_e$ \ci{13}. For this one must find such sets of partial parameters
$(\{\sigma'\},\{x'\})$,
which are invariant under the combined action of some inversion transformation
$I_{\sigma_0}$ and a set of some permutations $P$, i.e. the fixed points  of all
possible such combined actions. But, in contrast with a binary case $N=2,$
this cannot be done for arbitrary partial conductivities $\sigma_{i},$ because,
in general, a parameter $\sigma_0,$ for which $I_{\sigma_0}$ interchanges
the set of all $\sigma_{i}$ into itself for arbitrary $\sigma_{i},$ does not exist.
Such parameter can exist only for some special values of $\sigma_{i}.$
Then one can obtain an exact equation for $\sigma_e$ at
some special values of its arguments by the substitution of this
combination of transformations in the duality or inversion relations. The exact
equation will have a form
$$
\sigma_e^2(\{\sigma'\},\{x'\}) = \sigma_0^2,
\en(14)
$$
here $\sigma_0$ is a parameter of the involved inversion transformation.
The corresponding exact equation for function $f$ has the form
$$
 f^2(\{z'\},\{x'\})=1,
\en(14a)
$$
i.e. the exact value of $f$ at any fixed point is equal
to 1. Consequently, the possible exact values for $\sigma_e$ at these fixed
points must satisfy the next equalities
$$
\sigma_e(\{\sigma'\},\{x'\}) = \sigma_0 = (s_N)^{1/N}.
\en(15)
$$
If the inversion is the interchanging transformation $I_{ij}$ (A2), then
$$
\sigma_e = \sigma_0 = \sqrt{\sigma_i\sigma_j} = (s_N)^{1/N}, \quad i \ne j.
\en(15a)
$$
Thus, (15,15a) give a general form of the exact values of $\sigma_e$ in
these fixed points. The last equalities in (15) and (15a) give a strong
constraint on possible fixed point values of $\sigma_i.$

\bs

1. 3-phase case at arbitrary concentrations.

\bs

In order to find out possible fixed points let us consider first the case $N=3.$
Remember that in case of $N=2$ such fixed point exists for arbitrary $\sigma_i \; (i=1,2),$
but only at the equal phase concentrations $x_1=x_2=1/2$, when the interchanging
inversion transformation $I_{12}$ can be compensated by the permutation of the pairs
of arguments. This gives the known Keller -- Dykhne result (3).
The similar fixed point in case of $N=3$ was found by Dykhne at equal phase
concentrations $x_1=x_2$ and the additional constraint on $\sigma_3$
$$
\sigma_3^2 = \sigma_0^2 = \sigma_1 \sigma_2,
\en(16)
$$
which ensures a conservation of $\sigma_3$ under the inversion.
The corresponding exact value for $\sigma_e$ coincides with (3).

\begin{figure}
\begin{picture}(250,120)
\put(50,30){\vector(1,0){90}}
\put(50,30){\vector(0,1){90}}
\qbezier(130,40)(60,40)(60,110)
\put(145,30){$\sigma_1$}
\put(55,120){$\sigma_2$}
\put(75,5){(a)}
\put(120,0){\begin{picture}(100,50)%
\put(170,50){\vector(1,0){20}}
\put(120,100){\vector(0,1){20}}
\put(100,30){\vector(-1,-1){20}}
\put(100,30){\line(1,1){45}}
\put(170,50){\line(-3,1){60}}
\put(120,100){\line(1,-3){20}}
\put(170,50){\line(-1,1){50}}
\put(170,50){\line(-3,-1){75}}
\put(120,100){\line(-1,-3){25}}
\put(195,50){$x_2$}
\put(125,120){$x_3$}
\put(70,15){$x_1$}
\put(170,40){$1$}
\put(110,100){$1$}
\put(85,25){$1$}
\put(130,5){(b)}
\end{picture}}
\end{picture}

\vs{0.5cm}

{\small Fig.1. (a) A schematical plot of possible values of $\sigma_i, \; (i=1,2),$
satisfying the relation $\sigma_1\sigma_2 = \sigma_0^2$ with some fixed
value $\sigma_0$;
(b)  a concentration phase diagram for $N=3$, the bisectrisses correspond
to the pairwise equal phase concentrations.}
\end{figure}

Now we note that the Dykhne result (16) for $N=3$ admits a
generalization on the case of equal concentrations only of two first phases
$x_1=x_2=x,$ where $x$ can be an arbitrary concentration, satisfying
only the normalization condition $2x + x_3 = 1$. In this case, due to (16),
the pair of arguments $(x_3, \sigma_3)$ remains unchanged under
the inversion
transformation $I_{12}$ and the permutation of two first pairs of arguments.
Thus, the corresponding exact value for $\sigma_e$ coincides again with (16).
Of course, one can choose  the pairs with equal concentrations by various ways.
For example, if  the phases $1$ and $3$ have the equal concentrations
$x$ and $\sigma_0^2 = \sigma_1 \sigma_3 = \sigma_2^2,$ then one obtains
$$
\sigma_{e} = \sqrt{\sigma_1 \sigma_3} = \sigma_2.
\en(17)
$$
Analogously, when $x_2 = x_3$ and
$\sigma_0^2 = \sigma_2 \sigma_3 = \sigma_1^2,$ one obtains
$$
\sigma_{e} = \sqrt{\sigma_2 \sigma_3} = \sigma_1.
\en(18)
$$
The corresponding lines of equal concentrations of different phases
on the concentration phase diagram are shown on Fig.1b,
where they coincide with the bisectrisses of the triangle.
Their unique intersection point corresponds to the case of equal
concentrations of all phases.
In the $\sigma$-space the FP conditions (16)-(18) determine three surfaces.
Their intersections with a plane, defined by the condition
$\sum_1^3 \sigma_i = \sigma_c,$ where $\sigma_c$ is some constant,
are shown on Fig.2. Here all curves also intersect in one point, corresponding
to the equal partial conductivities.

\begin{figure}
\begin{picture}(250,120)
\put(80,0){\begin{picture}(100,50)
\put(120,50){\vector(1,0){20}}
\put(70,100){\vector(0,1){20}}
\put(50,30){\vector(-1,-1){20}}
\qbezier(120,50)(67,39)(70,100)
\qbezier(50,30)(75,73)(120,50)
\qbezier(70,100)(100,45)(50,30)
\qbezier(120,50)(95,75)(70,100)
\qbezier(50,30)(85,40)(120,50)
\qbezier(70,100)(60,65)(50,30)
\put(145,50){$\sigma_2$}
\put(75,120){$\sigma_3$}
\put(20,15){$\sigma_1$}
\put(120,40){$\sigma_c$}
\put(55,100){$\sigma_c$}
\put(35,30){$\sigma_c$}
\end{picture}}
\end{picture}

\vs{0.5cm}

{\small Fig.2. A schematical plot of
the intersection of a plane defined by
the relation $\sum_1^3 \sigma_i = \sigma_c,$ where
$\sigma_c$ is some constant, with the fixed point surfaces
$\sigma_i \sigma_j = \sigma_k^2,\; i \ne j \ne k, \; i,j,k=1,2,3,$
in the $\sigma$-space.}
\end{figure}

It follows from above results that $\sigma_e$ remains fixed, when $\sigma_i$ belong
to one of these surfaces and the concentrations
of the phases change along the corresponding lines!
It is worthwhile to note that now exact value of $\sigma_e$ at all equal
phase concentrations $x_1 = x_2 = x_3 =1/3$ is not universal, contrary
to the $N=2$ case, and depends on the constraint put on the $\sigma_i.$
The form of these possible exact values as a square root of product of two
partial conductivities or a value of the conductivity of the third phase is
determined by the form  of the interchanging inversion transformation and
by the structure of the corresponding fixed points.

\bs

2. General $N$-phase case.
\bs

We show now that the DR (8) for $N$-phase systems
admits a generalization of these results on the $N$-phase systems.

Of course, for systems with $N > 3$ there are similar  fixed points,
when $x_i = x_j$ and all other $\sigma_l \; (l\ne i,j)$ are equal between
themselves. But this case is trivial, since it reduces to the 3-phase case.
Fortunately, the new possibilities appear for $N>3.$

Let us consider firstly a case when $N$ is even $N=2M.$
Then a new fixed point is possible, which
corresponds to $M$ pairs of phases with equal concentrations
and equal products of the corresponding conductivities. For example,
the fixed point
$$
x_{2i-1}=x_{2i}=y_i, \quad
\sigma_0^2 = \sigma_{2i-1} \sigma_{2i} \quad (i=1,...,M)
\en(19)
$$
is possible. But all $\sigma_i$ in the phase pairs must be different.
In other words the conductivities of these pairs correspond to $M$
different  points and their inversed partners on the curve (see Fig.1a)
$$
\sigma_0^2 = \sigma_{1} \sigma_{2}.
$$
The effective conductivity has in this case the next exact value
$$
\sigma_{e} = \sigma_0 = \sqrt{\sigma_{2i-1} \sigma_{2i}} \quad
(i=1,...,M).
\en(20)
$$
Note that the equal concentrations of the phase pairs $y_a \; (a=1,...,M)$
can be arbitrary, except the normalizing condition $\sum_{a=1}^M 2y_a =1.$
The exact value (20) does not depend on $y_a$ and ensures again the correct
values for $\sigma_e$ in the limits $y_a \to 0, \; 1 \; (a=1,...,M).$
Of course, the similar fixed points exist for other ways of the partition
of $N$ phases into the pairs  with equal concentrations.
A number of such points is equal $\#_{2M} = (2M-1)!!.$
The exact values of $\sigma_e$ in these points have always the same general
form
$$
\sigma_{e} = \sigma_0 = \sqrt{\sigma_{i} \sigma_{j}} \quad
(i<j, \; i,j = 1,...,N),
\en(21)
$$
where possible pairs $(ij)$ correspond to the phases with equal concentrations
in the corresponding fixed points.

Now we consider a case when $N$ is odd $N=2M+1.$ In this case a new fixed
point is possible if the above conditions of the even case (19) are
supplemented by a condition on $\sigma_{2M+1}$ analogous to the Dykhne
condition for $N=3$ case
$$
\sigma_{2M+1}^2	= \sigma_0^2 = \sigma_{2i-1} \sigma_{2i} \quad (i=1,...,M).
\en(22)
$$
The corresponding effective conductivity  has again the exact value (20),
coinciding with the exact value of system with $N=2M$.
Of course, the other fixed points related with various ways of choice of
the phase pairs with equal concentrations are possible.
Their number is equal to
$\#_{2M+1} = (2M+1)\#_{2M} = (2M+1)!!.$
A general form of the exact values of $\sigma_e$ remains the same as in the
even case with an additional equality
$$
\sigma_e = \sigma_0 = \sigma_{k} = \sqrt{\sigma_{i} \sigma_{j}}, \quad
i\ne j \ne k, \;\; i <j \quad (i,j,k = 1,...,N).
\en(23)
$$
where the pairs $ij$ correspond to the phases with equal concentrations
and $k$ corresponds to the unpaired phases.

It follows from (19)-(23) that the FP conditions determine now  in the
$x$-space $M-1$-dimensional
hyperplanes for even $N=2M$ ($M$-dimensional hyperplane for odd $N=2M+1$)
and $M$-dimensional ($M+1$-dimensional) hypersurfaces in the $\sigma$-space.
Again $\sigma_e$ remains constant, when $\sigma_i$ belong to these hypersurfaces
and the concentrations $x_i$ change along the corresponding hyperplanes.
One can see from (21),(23) that the exact values of $\sigma_e$ at the point of
all equal concentrations $x_i=1/N$ are also nonuniversal.

It can be shown that only the fixed points of type (19),(22) and  their various
permutations are admissible for dual transformations of the form
$I_{\sigma_0}\prod P_b,$ when the action of the  inversion transformation
is equal to (or compensated by) the product of permutations
$P=\prod P_b$ (see Appendix B).

\bs

\cl{\bf 4. EM approximation for random $N$-phase systems}

\bs

Now, let us consider the EM approximation for the self-dual $N$-phase square lattice
RRN model (we assume it also as the EM approximation for all self-dual $N$-phase systems).
We will show that it has a very complicated form, and even to check a validity of
the  obtained exact results in it is not a simple task.
The EM approximation gives the following equation
for $\sigma_{e}$  \ci{2,3,4}
$$
\sum_1^N \; x_i \;\fr{\sigma_{e} - \sigma_i}{\sigma_{e} + \sigma_i} = 0.
\en(24)
$$
It can be written as an algebraic equation of the $N$-th order
$$
W(\sigma_{e}, \sigma_1,...,\sigma_N, x_1,...,x_N) =
\sum_{i=1}^N x_i (\sigma_{e} - \sigma_i)
\prodp_{k=1}^N (\sigma_{e} + \sigma_k) =
$$
$$
\sigma_{e}^N + a_1 \sigma_{e}^{N-1} + ... + a_{N-1} \sigma_{e} + a_N = 0,
\en(25)
$$
$$
a_k = s_k - 2 \bar s_k = -s_k +2\tilde s_k \quad  (k=1,...,N-1), \quad
a_N = - s_N.
\en(26)
$$
Here
$$
\bar s_k = \sum_{i=1}^N x_i \sigma_i
\sump_{i_1<i_2<...<i_{k-1}}^N \prod_{i_l=i_1}^{i_{k-1}} \sigma_{i_l}, \quad
\tilde s_k = \sum_{i=1}^N x_i \sump_{i_1<i_2<...<i_k}^N
\prod_{i_l=i_1}^{i_k}\sigma_{i_l},
\en(27)
$$
where all $\sump_{i_1<i_2<...<i_k}^N$ do not contain $i_l = i,$ and for $k=N$
$\bar s_N = s_N, \; \tilde s_N = 0.$
The coefficients $a_k$ are the homogeneous functions of $\sigma_i$
of degree k and the linear functions of concentrations $x_i.$
For this reason they are related
only with the corresponding symmetrical functions $s_k$ and their "averages"
$\bar s_k$ or $\tilde s_k$ (since there are different ways to define averaged
$s_k$).
The eq. (25) must have one physical solution $\sigma_e \ge 0,$
which depends on all symmetrical functions and their "averages".

It is easy to check
that $\sigma_{e}$, determined by the equation (25), satisfies (8a), since
the equation (24) for inversed partial conductivities has the next property
(for all $\sigma_i \ne 0$)
$$
\sum_1^N x_i
\fr{\sigma_{e} -\sigma_0^2/ \sigma_i}{\sigma_{e} + \sigma_0^2/\sigma_i} = -
\sum_{i=1}^N x_i
\fr{\sigma_0^2/\sigma_{e} - \sigma_i}{\sigma_0^2/\sigma_{e} + \sigma_i}=0.
\en(28)
$$
Moreover, since $a_N = - s_N$ one can pass directly to
the similar equation for $f$ by dividing all its terms on $s_N$
$$
f^N + \hat a_1 f^{N-1} + ... + \hat a_{N-1} f - 1 = 0,
\en(29)
$$
where now the coefficients are expressed through projective variables $z_i$
$$
\hat a_i = a_i(\{z\}) = a_i/(s_N)^{i/N}, \quad i=1,...,N-1.
\en(30)
$$
It is possible to write out the physical solution of (29) in an explicit
form only for $N=3,4,$ but even in $3$-phase case it has a complicated form,
which doesn't permit to investigate it in detail or to use it for description of any
experimental or numerical results (see Appendix C, where it is written out for a
completeness).

Here we will show that the EM approximation reproduces the exact values from section~3.
Firstly we consider the case of equal concentrations $x_i= 1/N.$
Then one can see that the coefficients $\hat a_k$  and
$f$ must depend only on symmetrical functions $\hat s_k \; (k=1,...,N-1)$
$$
\hat a_k =\frac{N-2k}{N} \hat s_k, \quad f=f(\hat s_1,..., \hat s_{N-1}),
\en(31)
$$
Due to the transformation rules of $\hat s_k$ (A10) the transformed $f$ in the
duality relation (12) has the form
$$
I f(\hat s_1,..., \hat s_{N-1}) = f(\hat s_{N-1},...,\hat s_1).
\en(32)
$$
Thus, one must have at the fixed point the next equalities
$$
\hat s_{N-k} = \hat s_k, \quad k=1,...,N-1.
\en(33)
$$
It is easy to check that they are satisfied at the fixed points (19,22) and
that the coefficients $\hat a_k$ satisfy at these points the equations
$$
\hat a_{N-k} = - \hat a_{k}.
\en(34)
$$
It follows from (34) that for even $N=2M$ the coefficient $\hat a_M =0.$
Consequently, one
has from (29),(34) that the polynomial in (29) can be represented in the
factorized form
$$
(f-1)P_{N-1}(f) = 0.
\en(35)
$$
where a polynomial $P_{N-1}(f)$ of degree $N-1$  has the next form
$$
P_{N-1}(f) = \sum_{i=1}^M f^{2M-i} \sum_{l=0}^{i-1}\hat a_l  +
f^{M-1} \sum_{l=0}^{M-1}\hat a_l  + (\sum_{l=0}^{M-2}f^l)
(\sum_{l=0}^{M-2}\hat a_l), \quad N=2M;
$$
$$
P_{N-1}(f) = \sum_{i=0}^M f^{2M-i} \sum_{l=0}^{i}\hat a_l +
(\sum_{l=0}^{M-1}f^l)(\sum_{l=0}^{M-1}\hat a_l), \quad N=2M+1.
\en(36)
$$
Since all coefficients of $P_{N-1}(f)$ are positive, one can see from (36)
that equations (29) and (35) have always only
one physical solution at any fixed point (19,22)
$$
f=1, \quad \sigma_e = (s_N)^{1/N} = \sqrt{\sigma_i \sigma_j} \quad (i\ne j).
\en(37)
$$
which coincides with the corresponding exact value from (15),(15a) or (20),(21).

Now we pass to the case of the FP with arbitrary pairwise equal concentrations (19),(22).
Firstly, we note that,
due to independence of coefficients $\hat a_k,$  one can write any solution
$f$ of (29) as a function of $\hat a_k$
$$
f= f(\hat a_1, ..., \hat a_{N-1})
\en(38)
$$
Let us consider how the coefficients $\hat a_k \; (k=1,...,N-1)$ transform  under
inversion transformations. As it is shown in the Appendix A the transformation rules
for $\hat a_k$ at arbitrary concentrations are
$$
I_{\sigma_0} \hat a_k = - \hat a_{N-k}, \quad (k=1,...,N-1).
\en(39)
$$
It follows from (39) that at any FP $(\{z'\},\{x'\})$ one must have again
$$
\hat a_k = - \hat a_{N-k}, \quad (k=1,...,N-1),
\en(40)
$$
what  reduces the polynomial in equation (29)   to the
factorized form
$$
(f-1)P_{N-1}(f,\{x'\}) = 0,
\en(41)
$$
where now the polynomial $P_{N-1}(f,\{x'\})$  depends on concentrations $\{x'\}.$
Due to this factorization, eqs.(25) have at the FP (19),(22) the required
solutions of the form (21),(23), which are the reduction  of the corresponding
exact values (15) at these points.

Thus, we see that in the EM approximation $\sigma_{e}$
of self-dual $N$-phase system is a solution of the
$N$-th order equation, which is an algebraic, homogeneous, a degree 1,
function of the set of $2N-1$
independent arguments
$\{\sigma\} = (\sigma_1,...,\sigma_N)$ and $\{x\} = (x_1,...,x_N),$
($\sum_1^N x_i =1$)
symmetric under the permutation of pairs of arguments $(\sigma_i, x_i)$
and simultaneously satisfying the inversion relation (8a) and the exact values (23).
The solving of the equation (25) in general $N$-phase case is an interesting
mathematical task due to its relation with the symmetrical and inversion groups.
Unfortunately, it is not so useful for the understanding of physics
of the problem, since already in 3-phase case the EM approximation gives
for $\sigma_{e}$ a complicated algebraic expression,
which is cumbersome and practically untractable (see Appendix C).
This can be explained by the fact that this solution depends on all
symmetrical functions $\bar s_i, \; (i=1,2,...,N).$
From the other side, it was shown in the binary case $N=2,$ that there are
other simple solutions of the DR, which, though  differ from the EM approximation
expression, but also satisfy  all necessary requierements and can describe $\sigma_e$
of some self-dual continuous inhomogeneous systems for not too large
inhomogeneities with the same accuracy as the EM approximation \ci{11}.
In the next sections we will show that analogous solutions and systems exist for
arbitrary $N$ and that they have the simple forms, convenient for approximate
description of the effective conductivity in a wide range of parameters.
Their simplicity can be explained by the next circumstence: they are related only with
one symmetrical function and its averaged versions.

\bs

\cl{\bf 5. Solutions of inversion relation}

\bs

1. General solution.

\bs
In this section we consider the possible functional forms compatible with the
duality (8) and/or inversion (8a) relations. Due to its generality, the DR cannot
define the effective conductivity unambiguously and completely. The diversity of
the possible functional forms of $\sigma_e$ can reflect a variety of all possible
inhomogeneity structures of random inhomogeneous systems. For this reason
it is very desirable to find all possible functional forms compatible with
the duality relation or to propose  a general method of their construction.
Since (8) and (8a) take a place for all self-reciprocal systems,
the functional forms compatible with the DR will be applicable
for all self-reciprocal systems. The different self-reciprocal systems are distinguished
by corresponding additional functional properties.
The self-dual case corresponds to the additional pairwise permutation symmetry
of $\sigma_e.$ The duality and/or inversion relations (8),(8a)
must be supplemented by the boundary conditions 1) - 3) enumerated above in the
Section 2, which turn out rather restrictive.

For further notation simplification it will be convenient to consider the DR in the
form (8b). Then, taking into account the properties of the inversion transformation (A3),
one can write out formally a general solution of (8b) in the following  form
$$
\sigma_{e}(\{\sigma\},\{x\}) = g(\{\sigma\},\{x\})/
g(\{1/\sigma\},\{x\}),
\en(42)
$$
where $g$ is an arbitrary function. But, one needs to remember that $\sigma_e$
must be a homogeneous function of a degree 1 and obey additional requirements.
The simplest way to satisfy the homogeneity
condition is to require $g$ to be also a homogeneous function. Then, if $g$ is
homogeneous of a degree $k,$ one can show that $\sigma_e$  is
$$
\sigma_{e}(\{\sigma\},\{x\}) = \left(g(\{\sigma\},\{x\})/
g(\{1/\sigma\},\{x\})\right)^{1/2k}.
\en(43)
$$
In order to obtain some $\sigma_e$ for
self-dual systems a function $g$ must be a pairwise
symmetrical function and satisfies the boundary conditions 1)-3).

It is useful to consider a general solution
in the projective variables $\{z\}.$ Then for a function $f$ from (11) one
obtains from (43) the general solution
$$
f(\{z\},\{x\}) = \left(g(\{z\},\{x\})/
g(\{1/z\},\{x\})\right)^{1/2k},
\en(44)
$$
where a function $g$ is a pairwise symmetrical and is homogeneous function
of a degree $k$ of $\{z\}.$
\bs

2. Some basic solutions.

\bs
Now we consider some simple solutions of the inversion relation for self-dual systems.

1) As a simplest admissible pairwise symmetrical function one can take the
mean conductivity $g(\{\sigma\},\{x\})= \langle \sigma \rangle = \bar s_1 =
\sum_{i=1}^N x_i \sigma_i,$ which
is a pairwise symmetrical homogeneous function of a degree 1. It satisfies
also the boundary conditions. The corresponding
function of the inversed partial conductivities $g(\{1/\sigma\},\{x\}) =
\langle \sigma^{-1} \rangle = \sum_1^N x_i/\sigma_i = \tilde s_{N-1}/s_N.$
Then, substituting these functions in (46) one obtains
$$
\sigma_{e}^{(1)}(\{\sigma\},\{x\}) = \sqrt{\langle \sigma \rangle
\langle \sigma^{-1} \rangle^{-1}} = \left(\bar s_1 s_N /\tilde s_{N-1}\right)^{1/2}
= (s_N)^{1/N} \left(\bar {\hat s}_1/\tilde {\hat s}_{N-1}\right)^{1/2}
$$
$$
= \left(\fr{ \sum_{i=1}^N x_i \sigma_i \prod_{k=1}^N \sigma_k}
{\sum_{i=1}^N x_i \prodp_{k=1}^N \sigma_k}\right)^{1/2}.
\en(45)
$$
It is clear from the form of the solution (47) that being a geometric average of
$\langle \sigma \rangle$ and $\langle \sigma^{-1} \rangle^{-1}$ it satisfies the
inequalities (5). It is easy to check that (47) satisfies the exact values at the FP.

2) Analogously, one can take as $g$  the averaged $k$-th moments
of a conductivity $\langle \sigma^k \rangle \; (k = 2,3,...),$
$$
g_k(\{\sigma\},\{x\}) = \langle \sigma^k \rangle = \sum_1^N x_i \sigma_i^k,
\en(46)
$$
which are the pairwise symmetrical homogeneous functions of a degree $k,$
satisfying boundary conditions.
They can be considered as the "averaged" version $\ol N^{(k)}_1$ of the newtonian
symmetrical functions $N_k = \sum_1^N \sigma_i^k.$
The corresponding  functions of the inversed partial conductivities are
$$
g_k(\{1/\sigma\},\{x\})= \sum_1^N x_i/\sigma_i^k =
\sum_1^N x_i \prodp_1^N \sigma_l^k/\prod_1^N \sigma_l^k =
\widetilde {N}_{N-1}^{(k)}/s_N^k,
\en(47)
$$
where $\widetilde {N}_{N-1}^{(k)} = \sum_1^N x_i \prodp_1^N \sigma_l^k.$
Substituting (46) and (47) into (43) one obtains the
solutions
$$
\sigma_{e}^{(k)}(\{\sigma\},\{x\}) = \left(\ol {N}_1^{(k)} s_N^k/
\tilde {N}_{N-1}^{(k)}\right)^{1/2k} =
(s_N)^{1/N}\left(\overline {\hat N}_1^{(k)}/\widetilde {\hat N}_{N-k} \right)^{1/2k}.
\en(48)
$$
These simple algebraic solutions can be considered as the basic solutions,
corresponding to the $k$-th moments of a conductivity.  They satisfy all necessary
requirements. It is worth to note here that the higher moments of conductivity
fluctuations are already accounted, due to the DR, even in the simplest solution (45).

In order to see the importance of the boundary condition requirements one can
consider as the functions $g_k$ the "averaged"
functions $\bar s_k$ or $\tilde s_k \; (k = 2,3,..., N-1),$ appearing in the EM
approximation in the section 4 and defined by (27). They
are homogeneous, pairwise symmetrical and, in some sense (see (A7)), inversed to each
other. Substituting these functions into (43) one obtains the following
solutions
$$
\sigma_{e}^{(k)}(\{\sigma\},\{x\}) =
\left( \bar s_k s_N/\tilde s_{N-k} \right)^{1/2k} =
$$
$$
(s_N)^{1/N}\left(\bar {\hat s}_k/\widetilde {\hat s}_{N-k} \right)^{1/2k},
\quad (k=2,...,N-1)
\en(49)
$$
or
$$
\sigma_{e}^{(k)}(\{\sigma\},\{x\}) =
\left(\ti s_k s_N/\bar s_{N-k} \right)^{1/2k} =
$$
$$
(s_N)^{1/N}\left(\wti {\hat s}_k/\ol {\hat s}_{N-k} \right)^{1/2k},
\quad (k=1,...,N-2).
\en(50)
$$
Unfortunately, these solutions do not satisfy the reduction and boundary conditions
from the Section 2, because, for example, for $\sigma_1= \sigma_2$
$$
\bar s_k^{(N)} \to \bar s_k^{(N-1)} + \sigma_1 \bar s_{k-1}^{(N-1)}, \quad
\ti s_k^{(N)} \to \ti s_k^{(N-1)} + \sigma_1 \ti s_{k-1}^{(N-1)},
$$
where the upperscript $(N)$ denotes the corresponding $N$-phase system. It
follows from these formulas that under the reduction $\sigma_1= \sigma_2$
the functions $\bar s_k^{(N)}$ and $\ti s_k^{(N)}$ (except $\bar s_1$ and
$\ti s_{N-1}$) relate with the corresponding lower functions.
For this reason the solutions (51) and (52) are not physical ones.
Since the "pure" symmetrical functions reduce by the same law, the coefficients
functions $a_k$ from the section 4 also reduce for $\sigma_1= \sigma_2$ by the same law,
and because the solutions connected with them are also nonphysical ones.

3) It is not possible to define in this way "an average" for $N$-th
symmetrical function. Fortunately,
there is a simple pairwise symmetrical exponential function, generalizing
the $N$-th symmetrical function, the monomial
$\prod_{i=1}^N \sigma_i^{x_i}.$ It can be represented as an exponential of
the simplest pairwise symmetrical function
$$
g = \exp \sum_{i=1}^N b_i x_i, \quad b_i=\ln \sigma_i.
$$
It is easy to see that it is invariant under the joint
inversion  of the partial conductivities and of the function itself
$$
g^{-1}(\{1/\sigma\},\{x\}) = g(\{\sigma\},\{x\}).
$$
For this reason its substitution into (43) reproduces itself
$$
\sigma_{e}^{(N)}(\{\sigma\}, \{x\}) =
\prod_{i=1}^N \sigma_i^{x_i}.
\en(51)
$$

The solutions (45) and (51) generalize two solutions found earlier for 2-phase systems
\ci{9,12}. Since they satisfy the inversion relation, they automatically
have all necessary properties determined by this relation. For example,
it is easy to check that (51) reproduces at the corresponding fixed points
the  exact values from (20),(21),(23) due to the normalization condition for
concentrations
$$
\sigma_e = \prod_{i=1}^M \sigma_0^{2y_i} = \sigma_0 = (s_N)^{1/N},
\quad (N=2M),
\en(52)
$$
$$
\sigma_e = \sigma_0^{1-2\sum_1^M y_i} \prod_{i=1}^M \sigma_0^{2y_i} =
\sigma_0 = (s_N)^{1/N}, \quad (N=2M+1).
\en(53)
$$
It is interesting that in this case $\sigma_e$ at the point of all equal
concentrations $x_i=1/N$ always coincides with $s_N^{1/N},$ (i.e. for arbitrary
$\sigma_i$).
The solution (51) satisfies the inequalities (5) due to the H\"older
inequalities
$$
\prod_1^N \sigma_i^{x_i} \le \sum_1^N x_i \sigma_i, \quad
\prod_1^N \sigma_i^{-x_i} \le \sum_1^N x_i/\sigma_i, \quad
$$
$$
\sum_1^N x_i = 1, \quad \sigma_i > 0.
\en(54)
$$

The physical models with effective conductivities equal to the simplest solutions
(45) and (51) will be represented in the two next sections, where the other physical
properties of these solutions will be also considered.
The basic solutions (48) can be used also as constructive blocks for more sophisticated
solutions. For example, let $f_i \; (i=1,2,...n)$ are the solutions, satisfying all
requirements. Then
$f=\left(\prod_1^n f_i \right)^{1/n}$ is also the solution, satisfying all requirements.
The physical models related with the higher moments solutions (48) remain unknown at this
time. One can also use as constructive blocks the solutions (49), (50), but this is a
complicate task to arrange them in a combination, satisfying all necessary requirements.
The EM approximation case confirms this.
\bs

\cl{\bf 6. FMSA approximation and functional equation}

\bs

In this section we show that the solution (51) can describe $\sigma_e$ of continuous
inhomogeneous self-dual systems with compact inclusions of different phases.
A schematical picture of such systems is shown on Fig.3a.
In order to do this we will generalize an approximate functional equation
for $\sigma_{e}$ of two-phase systems from \ci{12} on $N$-phase systems and
solve it.

To obtain this functional equation we can use a method called the FMSA
approximation and proposed in \ci{12}.
One can see that all arguments represented there in favor of
the FMSA approximation hold also for random inhomogeneous heterophase systems
with any finite number of phases $N \ge 3$ .


\begin{figure}
\begin{picture}(250,120)
\put(45,20){\line(0,1){50}}
\put(45,20){\line(1,0){50}}
\put(95,20){\line(0,1){50}}
\put(45,70){\line(1,0){50}}
\put(50,25){$\circledast$}
\put(85,40){$\circledast$}
\put(55,60){$\circledast$}
\put(67,50){$\circledast$}
\put(49,60){$\circledcirc$}
\put(70,55){$\circledcirc$}
\put(79,25){$\circledcirc$}
\put(65,45){$\circledast$}
\put(75,40){$\circledast$}
\put(55,30){\circle*{13}}
\put(60,55){\circle*{7}}
\put(75,60){\circle*{10}}
\put(85,40){\circle*{10}}
\put(65,45){\circle*{5}}
\put(85,35){\circle*{5}}
\put(80,55){\circle*{5}}
\put(54,35){\circle*{5}}
\put(60,35){\circle*{5}}
\put(67,43){\circle*{5}}
\put(83,25){\circle*{5}}
\put(83,57){\circle*{5}}
\put(48,40){$\Cup$}
\put(48,45){$\Cap$}
\put(68,30){$\Cap$}
\put(84,60){$\Cap$}
\put(84,55){$\Cup$}
\put(65,63){$\circledast$}
\put(68,25){$\circledast$}
\put(58,25){$\circledast$}
\put(65,8){({\small a})}
\put(130,0){%
\begin{picture}(100,50)%
\multiput(75,20)(10,0){11}%
{\line(0,1){50}}
\multiput(75,20)(0,10){6}%
{\line(1,0){100}}
\multiput(77.5,21)(0,20){3}{1}
\multiput(77.5,31)(0,20){2}{2}
\multiput(87.5,31)(0,20){2}{1}
\multiput(87.5,21)(0,20){3}{2}
\multiput(97.5,21)(0,20){3}{1}
\multiput(97.5,31)(0,20){2}{2}
\multiput(107.5,31)(0,10){3}{2}
\multiput(107.5,21)(0,40){2}{1}
\multiput(117.5,51)(0,10){2}{2}
\multiput(117.5,21)(0,10){3}{1}
\multiput(127.5,21)(0,10){5}{2}
\multiput(137.5,21)(0,20){3}{1}
\multiput(137.5,31)(0,20){2}{2}
\multiput(147.5,21)(0,10){5}{1}
\multiput(157.5,31)(0,10){2}{2}
\multiput(157.5,21)(0,30){2}{1}
\multiput(167.5,31)(0,20){2}{2}
\multiput(167.5,21)(0,20){2}{1}
\put(167.5,61){2}
\put(157.5,61){2}
\put(120,5){({\small b})}
\end{picture}}
\end{picture}

{\small Fig.3. a) A schematic picture of the elementary square with a size
$l \gg l_m$ of one phase with compact inclusions of the other phases,
b) the auxiliary lattice of the model, the numbers 1,2 denotes
squares with the corresponding sets of concentrations $\{x\}_{1,2}$.}
\end{figure}
To do this let us consider the auxiliary square lattice with the squares of a size
$l \gg l_{max},$  where $l_{max}$ is a higher boundary value of $l_m(x),$
the maximal size of inhomogeneities of the system (which can depend on the phase
concentrations), i.e. $l_m(x)$ always remains smaller than
$l_{max}.$ It is assumed that the initial inhomogeneous system has such $l_{max},$
at least, it exists in some range of the partial concentrations.
We require that the squares of the auxiliary lattice have the
effective conductivities corresponding to two different sets of the
concentrations $\{x\}_1$ and $\{x\}_2$ with equal probabilities $p = 1/2$ (see Fig.3b).
Then, implementing to this lattice the exact Keller-Dykhne formula (3)
and supposing for a self-consistency that $\sigma_e$ of this lattice
must be equal to the same effective conductivity at the averaged concentrations
$\{x\} = (\{x\}_1+\{x\}_2)/2,$
one can obtain the approximate functional equation for the effective conductivity
of this system as a function of all partial concentrations $\{x\}$
$$
\sigma_{e}(\{\sigma\},\{x\}) = \sqrt{\sigma_{e}(\{\sigma\},\{x\}_1)
\sigma_{e}(\{\sigma\},\{x\}_2)},
$$
$$
\{x\} = (\{x\}_1+\{x\}_2)/2, \quad (\sigma_i \ne 0).
\en(55)
$$
The eq.(55), in contrast with the DR, connects $\sigma_e$ at different concentrations,
but at the same conductivities.
Strictly speaking, the eq. (55) takes place only for the ranges of $x_i,$
where $l_{max}$ exists. These ranges can be enlarged by taking the larger $l.$
The boundary conditions for the equation (55) follow from the conditions
1) - 3) of the Section 2
$$
\sigma_e (\{\sigma\},x_i=1)= \sigma_i, \quad
\sigma_e (\{\sigma\},x_i=1)= \sigma_i, \quad
\en(55')
$$
Let us suppose for a moment that the equation (55) is applicable for all
phase concentrations.
Then a solution of (55) has an exponential form, generalizing  the solution of
two-phase systems from \ci{9,12}
$$
\sigma_{e}(\{\sigma\},\{x\}) = \prod_{i=1}^N {\sigma_i}^{x_i}, \quad
\sum_{i=1}^N x_i = 1.
\en(56)
$$
It  coincides exactly with the solution (45) and
has the simplest form compatible with both symmetry groups (symmetrical
and inversion ones) of the problem. We see that constrainting himself by the
RRN type models or by the EM approximation and the corresponding
algebraic equation one cannot obtain the more simpler exponential solution.
At equal concentrations of all phases $x_i = 1/N$ one obtains from (56)
$$
\sigma_{e}(\{\sigma\},\{x\}) = \prod_{i=1}^N {\sigma_i}^{1/N} = (s_N)^{1/N}.
\en(57)
$$
This formula generalizes the exact Keller -- Dykhne formula (3) for systems
with $N=2$ on the case of random inhomogeneous systems with arbitrary $N$
phases and the exact values (23) for $N$-phase systems  at the FP on all
$\sigma_i \ne 0.$

But, as it was mentioned in the section 3, the formulas for $\sigma_e$ at $x_i=1/N$
are not universal. For example, it differs from the corresponding formula in the
EM approximation (see Appendix C).
Below we will obtain  another formula for $\sigma_e$ of the random
$N$-phase model with other structure of inhomogeneities, which differs at $x_i=1/N$ from
both these formulas.

It is easy to check that in the weak inhomogeneous limit, when all partial
conductivities $\sigma_i$ can be represented in the form
$$
\sigma_i = \sigma_0(1+\delta \sigma_i),
$$
where $\sigma_0$ is a constant homogeneous conductivity, the formula (56)
gives up to the second order on $\delta \sigma_i$
$$
\sigma_e = \sigma_0 \prod_i (1+\delta \sigma_i)^{x_i} \approx
\sigma_0 \left[1+ \sum_1^N x_i \delta_i -\fr{1}{2}(\sum x_i \delta_i^2
+ (\sum x_i \delta_i)^2)\right]
$$
which reproduces the general Landau-Lifshitz
formula (13) for $\sigma_0 = \langle \sigma \rangle,$ when $\sum x_i \delta \sigma_i =0.$
The formula (56) means  that $\ln \sigma$ is a self-averaging quantity
in this approximation, because it can be represented in the form
$$
\sigma_{e}(\{x\},\{\sigma\}) = \exp (\sum_{i=1}^N x_i \ln \sigma_i) =
\exp \langle \ln \sigma \rangle = \exp \langle b \rangle.
\en(58)
$$
This property noted firstly by Dykhne for the two-phase systems at equal
concentrations $x=1/2$ \ci{11} was established later in the weak localization theory
of conductivity of two-dimensional disordered systems \ci{18}.
Since the solution (56) reproduces the exact values at all FP, it can be used in
a more wide range of concentrations than it was assumed under the deducing of
the eq.(55). We expect that it works well for $\sigma_i > 0$ at
{\it all concentrations} for all self-dual inhomogeneous systems, to which the FMSA
approximation is applicable. For example, it can be applyed to the inhomogeneous
disordered systems where the regions with inhomogeneously distributed impurities
can be described effectively as the regions of different phases.
One needs to note that the formula (53) for arbitrary partial concentrations
have been proposed earlier as a result of the extrapolation of the analogous formula
for $\sigma_e$ of some special models of random isotropic polycrystal systems
at discrete set of concentrations on arbitrary  concentrations \ci{19}.

\bs

\cl{\bf 7. Random parquet model of heterophase systems}

\bs

Now we will construct  a hierarchical model of two-dimensional isotropic randomly
inhomogeneous heterophase system, using the composite method introduced
in \ci{9,12}, and show that  its effective
conductivity $\sigma_{e}$ obtained in the mean field like approximation for arbitrary
values of the phase concentrations $x_i$ coincides with the solution (45) of the DR.

Let us consider the following two-dimensional model. There is a simple
square lattice with the squares consisting of a random layered(or striped) mixture
of $N$ conducting phases with constant conductivities $\sigma_i, i = 1,...,N$ and
the corresponding concentrations $x_i.$   A schematic picture of such
square is given in Fig.4.


\begin{figure}
\begin{picture}(250,120)
\put(50,20){\line(1,0){50}}
\put(50,20){\line(0,1){50}}
\put(100,20){\line(0,1){50}}
\put(50,70){\line(1,0){50}}
\put(57,20){\line(0,1){50}}
\put(60,20){\line(0,1){50}}
\put(88,20){\line(0,1){50}}
\put(58.5,20){\line(0,1){50}}
\put(86,20){\line(0,1){50}}
\put(90,20){\line(0,1){50}}
\multiput(62.5,45)(2.5,0){3}{\circle*{0.5}}
\multiput(78,45)(2.5,0){3}{\circle*{0.5}}
\multiput(58.5,21)(0,3){17}{\circle*{1.5}}
\multiput(86,16)(0,3){17}{*}
\multiput(67,21)(0,7){7}{$\blacksquare$}
\put(66,5){(a)}
\put(130,0){%
\begin{picture}(100,50)%
\multiput(75,20)(10,0){11}%
{\line(0,1){50}}
\multiput(75,20)(0,10){6}%
{\line(1,0){100}}
\put(120,5){(b)}
\multiput(78,25)(0,20){3}{\line(1,0){3}}
\multiput(80,33)(0,20){2}{\line(0,1){3}}
\multiput(88,35)(0,20){2}{\line(1,0){3}}
\multiput(90,23)(0,20){3}{\line(0,1){3}}
\multiput(98,25)(0,20){3}{\line(1,0){3}}
\multiput(100,33)(0,20){2}{\line(0,1){3}}
\multiput(110,33)(0,10){3}{\line(0,1){3}}
\multiput(108,25)(0,40){2}{\line(1,0){3}}
\multiput(120,53)(0,10){2}{\line(0,1){3}}
\multiput(118,25)(0,10){3}{\line(1,0){3}}
\multiput(130,23)(0,10){5}{\line(0,1){3}}
\multiput(138,25)(0,20){3}{\line(1,0){3}}
\multiput(140,33)(0,20){2}{\line(0,1){3}}
\multiput(148,25)(0,10){5}{\line(1,0){3}}
\multiput(160,33)(0,10){2}{\line(0,1){3}}
\multiput(158,25)(0,30){2}{\line(1,0){3}}
\multiput(170,33)(0,20){2}{\line(0,1){3}}
\multiput(168,25)(0,20){2}{\line(1,0){3}}
\put(170,63){\line(0,1){3}}
\put(160,63){\line(0,1){3}}
\end{picture}}
\end{picture}

\vs{0.5cm}
{\small Fig.4. (a) An elementary square of the model with a
vertical orientation of the layers of different phases;
(b)  a lattice of the model, the small lines on the
squares denote their orientations.}
\end{figure}
The layered structure
of the squares means that the squares have some preferred direction,
for example, along the layers.
Let us suppose that the directions of different squares
are randomly oriented (parallely or perpendicularly) relatively to the
external electric field, which is directed along $x$ axis.
In order for system to be isotropic the probabilities of the parallel and
perpendicular orientations of squares must be equal or (what is the same)
the concentrations of the squares with different orientations must be equal
$p_{||} = p_{\perp} = 1/2.$

Such lattice can model a random system consisting from mixed phase regions,
which can be roughly represented on the small macroscopic scales as randomly
distributed plots with the effective "parallel" and "serial" connections
of the layered or striped $N$-phase mixture (Fig.4). The lines on the squares denote
their orientations. This structure can appear,
for example, on the intermediate scales when a random medium is formed
as a result of the stirring of the $N$-phase mixture or as a result of spontaneous
generation of  randomly distributed
regions of $N$ striped phases with different orientations of stripes near the phase
transition point. For external electric field directed along one of the preferred
orientation the corresponding averaged parallel and
perpendicular conductivities of squares $\sigma_{||}(x)$ and
$\sigma_{\perp}(x)$ are defined by the following formulas
$$
\sigma_{||}(\{\sigma\},\{x\}) = \sum_1^N x_i \sigma_i =
\langle \sigma\rangle,
$$
$$
\sigma_{\perp}(\{\sigma\},\{x\}) =
\left(\sum_1^N x_i/\sigma_i \right)^{-1} =
\langle \sigma^{-1}\rangle^{-1}.
\en(60)
$$
Thus we have obtained the hierarchical representation of random medium
(in this case a two-level one). On the first level it consists  from
some regions (the squares with two possible stripe orientations) of the random
mixture of the $N$ layered conducting phases with different conductivities
$\sigma_i$ and arbitrary concentration. On the second level
this medium is represented as a random parquet constructed from two such
squares with different conductivities $\sigma_{||}$ and $\sigma_{\perp}$,
depending nontrivially on concentrations of the initial conducting phases,
and randomly distributed with the same probabilities $p_i = 1/2$ (Fig.4).
This representation allows us to divide the averaging process into two
steps. Firstly, we average over each square and obtain for them $\sigma_{||}$ or
$\sigma_{\perp}.$ Then, on the second step, the averaging is done over the lattice
of squares and the exact Keller - Dykhne formula (3) is used.
As a result one obtains for the effective conductivity of the N-phase random
parquet model the following formula, which is applicable for
arbitrary concentration
$$
\sigma_{e}(\{\sigma\},\{x\}) = \sqrt{\sigma_{||}\sigma_{\perp}} =
\sqrt{\langle \sigma\rangle \langle \sigma^{-1} \rangle^{-1}} =
\left(\fr{\prod_{k=1}^N \sigma_k \sum_{i=1}^N x_i \sigma_i}
{\sum_{i=1}^N x_i \prod_{k \ne i}^N \sigma_k}\right)^{1/2}.
\en(61)
$$
It coincides with the solution (45) from the section 5 and generalizes the corresponding
solution for two-phase random parquet model from \ci{9,12}.
It is worth to note here that such formula for $\sigma_{e}$ has been obtained earlier
for inhomogeneous media with a smooth $\log$-symmetrical distribution function,
when it is the exact one \ci{11}.
Since such inhomogeneous media correspond to the limit
$N \to \infty$ and a smooth $\log$-symmetrical distribution function of the
random parquet model, the solution (61) becomes exact in this limit.
Analogous expression for $\sigma_e$ has been also proposed for one model
of random isotropic polycrystal systems with a  structure of inhomogeneities
similar to those of the random parquet model \ci{19}.

One can also check that (61) reproduces the FP
values (15), (21) and (23).
In a case of the equal phase concentrations $x_i=1/N$ one obtains now
$$
\sigma_{e}(\{\sigma\},\{1/N\}) = \sqrt{s_1 s_N/s_{N-1}}.
\en(62)
$$
Thus we see that the expressions (45),(51) and the EM approximation (Appendix C) for
the effective conductivity have different functional forms.
This fact means that the effective conductivity of self-dual random inhomogeneous
systems is a nonuniversal function. Moreover, it follows from these expressions
that the values of $\sigma_e$ at equal phase concentrations $x_i = 1/N$ and
arbitrary $\sigma_i$ are also nonuniversal (unlike the case $N=2$).
This is a consequence of the nonuniversality of the effective
conductivity and a fact that this point at arbitrary partial conductivities
is not the fixed point.
A nonuniversality of $\sigma_e$ at $x_i=1/3$ for some 3-phase self-dual
{\it regular} inhomogeneous systems was also noted earlier on a basis of numerical calculations
in the paper \ci{16}.

One can easily find from the explicit formulas (45) and (51) the corresponding
low concentration asymptotics of $\sigma_e$ and see that their coefficients
diverge at $\sigma_i \to 0.$  This shows that the limits $\sigma_i \to 0$ can
have some singularities as in the two-phase case \ci{9,12}.

\bs

\cl{\bf 9. Conclusions}

\bs
Thus, using the duality relation and some additional assumptions, we found out
a general form of the effective conductivity $\sigma_e$ of two-dimensional
self-dual heterophase systems and written out some explicit approximate expressions
for $\sigma_e$  at arbitrary values of partial concentrations and conductivities. .
Though these formulas  have the different functional forms, they

(1) satisfy all necessary requirements including symmetries and
inequalities,

(2) reproduce the general Landau - Lifshitz formula for $\sigma_{e}$
in the weakly inhomogeneous case,

(3) reproduce all exact values for $\sigma_{e}$ at the fixed points.

One of them is also a solution of the approximate functional equation,
and can describe $\sigma_{e}$ of various self-dual (or close to them)
heterophase systems with a compact inclusions of phases or with a finite maximal
scale of inhomogeneities $l_{max}$.
We have constructed  also a random parquet model of
the random inhomogeneous medium of the layred type and have shown that its
effective conductivity in some mean field type approximation coincides with
the solution connected with a mean conductivity.
All these results show also that, in general, $\sigma_{e}$ of the
inhomogeneous heterophase self-dual systems may be a nonuniversal function and
can depend on some details of the structure of the randomly inhomogeneous
regions.
The obtained explicit formulas for $\sigma_{e}$ can be used for an approximate
description of the effective conductivity of some real random and regular
inhomogeneous systems.

\bs

\cl{\bf Appendix A. Inversion, symmetrical functions and fixed points}

\bs

 In this appendix we consider the inversion group and its action on
various symmetrical functions. The inversion group with a parameter $\sigma_0,$
acting on one variable function $f(\sigma)$, contains two elements
$$
I_{\sigma_0} f(\sigma) = f(\sigma_0^2/\sigma), \quad  I_{\sigma_0}^2 = 1.
\en(A1)
$$
Since the  effective conductivity of the $N$-phase system depends
on $N$ partial conductivities, one must consider the action of
the direct products of the partial inversion transformations
$I^{(i)}_{\sigma_0}: \sigma_i \to \sigma_0^2/\sigma_i, \; (i=1,2...,N).$
Each partial inversion transformation acts only on the partial
conductivity of the corresponding phase and is characterized by an arbitrary
parameter $\sigma_0.$ Formally, the inversion transformation is well defined
only for all $\sigma_i \ne 0.$ One can extrapolate it on the case, when some
$\sigma_i = 0.$ In this case it transforms a system with $\sigma_i = 0$  into
a system with  $\sigma_i = \infty.$ Note also that the inversion
transformation with parameter  $\sigma_0$  has only one fixed point
$\sigma_i = \sigma_0.$ The i-th and j-th partial inversion transformations
$I^{(i)}_{ij} \;(I^{(j)}_{ij})$, having
$\sigma_0^2 = \sigma_i \sigma_j,$ are very important, because they change
$\sigma_i$ into $\sigma_j$ and vice versa
$$
I^{(i)}_{ij} \sigma_i \equiv I^i_{ij} =
\sigma_i \sigma_j/\sigma_i =
\sigma_j,\;
I^{(j)}_{ij} \sigma_j = \sigma_i \sigma_j/\sigma_j =
\sigma_i.
\en(A2)
$$
The physical sence has only the product of all partial inversion
transformations with the same parameter $\sigma_0$
$$
I_{\sigma_0} = \prod_{i=1}^N \; I_{\sigma_0}^{(i)}, \quad  I_{\sigma_0}^2 = 1.
\en(A3)
$$
Since in our problem the inversion transformation acts on a space of
symmetrical homogeneous functions it is useful to consider its action on the
basic elementary (and newtonian) symmetrical functions
$s_k(\{\sigma\}) \; (k=1,...,N, \{\sigma\} = (\sigma_1,...,\sigma_N))$
$$
s_1 = \sum^N_{i=1} \sigma_i,\quad
s_2 = \sum^N_{i<j} \sigma_i \sigma_j, \quad
s_k = \sum^N_{i_1<i_2<...<i_k} \prod_{i_1}^{i_k} \sigma_{i_l}, \quad
$$
$$
I_{\sigma_0} s_k = \sigma_0^{2k} s_{N-k}/s_N.
\en(A4)
$$
It follows from (A4) that only $s_N$ transforms into itself. Under
consideration of the inversion transformation action on homogeneous functions
it is convenient to pass (for $\sigma_i \ne 0$) to the projective variables
$z_i$ and
introduce projective basic symmetrical functions $\hat s_k$
$$
\hat s_k = s_k(\{z\}) = s_k/(s_N)^{k/N}, \quad k=1,.., N-1.
\en(A5)
$$
Then we have from (A4) and (A5)
$$
I_{\sigma_0} z_i = 1/ z_i, \quad
I_{\sigma_0} \hat s_k = \hat s_{N-k},
\en(A6)
$$
i.e. the inversion transformation $I_{\sigma_0}$ for any $\sigma_0$
inverses $z_i$ and
interchanges adjoint $\hat s_k!$ This property will strongly simplify a
search of fixed points needed for obtaining the exact equations for
$\sigma_e.$
Now we consider how  the coefficients $a_k \; (k=1,...,N-1)$ from the EMA
equation (25) transform under  inversion transformations.
Since $a_k = s_k -2\bar s_k = -s_k + 2\tilde s_k,$ one needs to find the
transformation rules for $\bar s_k$ and/or  $\tilde s_k.$ Using the definitions
of $\bar s_k$ and $\tilde s_k$ from (29) and one can show that
$$
I_{\sigma_0} \bar s_k = \tilde s_{N-k}/s_N, \quad
I_{\sigma_0} \tilde s_k = \bar s_{N-k}/s_N,
\en(A7)
$$
In the projective variables $z_i$ these transformations simplify
$$
I_{\sigma_0} \bar {\hat s}_k =  \tilde {\hat s}_{N-k}, \quad
I_{\sigma_0}  \tilde {\hat s}_k =  \bar {\hat s}_{N-k},
\en(A8)
$$
These relations generalize the relations for pure symmetrical functions (A6) on a case
of arbitrary $x_i.$
It follows from (A7) and (A8)  that the transformation rules for $\hat a_k$ at arbitrary
concentrations are
$$
I_{\sigma_0} \hat a_k = - \hat a_{N-k}, \quad (k=1,...,N-1).
\en(A9)
$$
The fixed points $\{\sigma'\}$ of the inversion transformation $I_{\sigma_0}$
are defined as
$$
I_{\sigma_0} \{\sigma'\} = \{\sigma'\}
\en(A10)
$$
Thus, at any FP one must have
$$
\hat a_k (\{\sigma'\},\{x'\})= - \hat a_{N-k}(\{\sigma'\},\{x'\}), \quad (k=1,...,N-1).
\en(A11)
$$
\bs

\cl{\bf Appendix B. Evidence of absence of other fixed points}

\bs

One can show that only the fixed points of type (19,22) and  their various
permutations are admissible for all dual transformations of the form
$T(I,P) = \prod I_a \prod P_b,$ where, for clarity, both products are assumed
to be finite, and the action of the product of inversions
$I=\prod I_a$ must be equal to (or compensated by) the product of permutations
$P=\prod P_b$ (we will not consider here possible, more sophisticated, forms
of $T$).
To do this it is useful to pass from partial conductivities $\sigma_i$ to
their logarithms $b_i = \ln \sigma_i, \; (-\infty \le b_i \le \infty).$
Then the partial inversion transformation $I^{(i)}_{\sigma_0}$ acts on $b_i$ as a
reflection $b_i \to B - b_i, \; B = 2\ln \sigma_0.$ The inversion
$I_{\sigma_0}$ acts on the set of variables $\{b\}$ in the following way
$$
b_i \to B - b_i, \; (i=1,...,N)
\en(B1)
$$
The action of the product of $L$ different inversions
$I = \prod_{s=1}^L I_{\sigma_0^{(s)}}$ gives analogous result
$$
b_i \to B_L + (-1)^L b_i, \; (i=1,...,N)
\en(B2)
$$
with $B_L = \sum_{s=1}^L (-1)^s B_s, \; B_s = 2\ln \sigma_0^{(s)}.$
Since all parameters of inversion transformations collect into one constant
$B_L,$ it will be enough to consider only $L=1$ or $L=2.$
In order to find any fixed point the result () must be equivalent to some
permutation $P$ of the set $\{b\}$
$$\{b\}_P = (b_{1_P},...,b_{N_P}).$$
Let us consider
firstly a case of even $L=2.$ Then one obtains from (25)
$$
b_{i_P} = B_2 + b_i, \; (i=1,...,N)
\en(B3)
$$
The set of equations (B3) is very restrictive. If any $b_i$ remains unchanged,
than $B_2 = 0$ and all other $b_i$ also remain unchanged. It means that
in this case $I=1.$ If () gives a permutation of any  pair
$b_i \lra b_k, \; (i\ne k),$ than it follows from () that again $B_2 = 0.$
As is known any permutation $P$ decomposes into product of independent chains
of simplest permutations. Then,
considering all possible chains similar to the abovementioned simplest case
of pair permutation, one can see that equations (26) take place only when
$B_2=0.$ In case of odd $L=1$ equations (25) take the form
$$
b_{i_P} = B_1 - b_i, \; (i=1,...,N)
\en(B4)
$$
If (27) gives a permutation of some pair $b_i \lra b_k, \; (i\ne k),$ then
one obtains for $b_i,b_k$ only one equality (since in this case both equations
coincide)
$$
b_i + b_k = B_1.
\en(B5)
$$
In case of any larger chains of permutation one obtains additional
equalities, which have only trivial solution. For example, in case of
3-chain ($1\to 2 \to 3 \to 1$) they are
$$
b_1 + b_2  = b_2 + b_3 = b_1 + b_3 = B_1.
\en(B6)
$$
It follows from (29) that this is possible only if
$b_1=b_2=b_3 =b, \; B_1 =2b,$ but this corresponds to the one phase case.
Consequently, the equations (27) can take place if and only if
the permutation $P$ decomposes into product of pair permutations (for $N=2M$)
or into product of pair permutations and one identical transformation
(for $N=2M+1$).
Then one obtains in case of pair permutations of type $b_{2i-1} \lra b_{2i}$
the next equalities
$$
b_{2i-1} + b_{2i} = B_1 \quad (i=1,...,M), \quad (N=2M),
$$
$$
b_{2i-1} + b_{2i} = 2b_{2M+1} = B_1 \quad (i=1,...,M), \quad (N=2M+1).
\en(B7)
$$
All other solutions can be obtain from (30) by various permutations.
These solutions give exactly the fixed points (19,22)
after returning to the variables $\{\sigma \}.$
\bs

\cl{\bf Appendix C. EM approximation for the $3$-phase RRN model}

\bs

For 3-phase case the solution of the EMA equation can be presented in an explicit form,
but we will see that it is very complicated and inconvenient for a description of
experimental data and theoretical analysis.
According to the general results of the section 4 the equation for effective
conductivity of the random $3$-phase model is
$$
\sigma_e^3 + a_1 \sigma_e^2 + a_2 \sigma_e + a_3 =0, \quad
$$
$$
a_1=s_1-2\bar s_1 = -s_1 +2\tilde s_1,\; a_2=s_2-2\bar s_2 = -s_2 + 2\tilde s_2,\;
c=-s_3.
\en(C1)
$$
Here $\bar s_1=\bar \sigma = \sum_1^3 x_i \sigma_i$ is a mean conductivity,
$\tilde s_1 = x_1 (\sigma_2 + \sigma_3) + x_2 (\sigma_1 + \sigma_3) +
x_3 (\sigma_{1} + \sigma_{2}),$
$\bar s_2 = x_1 (\sigma_{12} + \sigma_{13}) + x_2 (\sigma_{12} + \sigma_{23}) +
x_3 (\sigma_{13} + \sigma_{23}),$
$\ti s_2 = x_1 \sigma_{23} + x_2 \sigma_{13} + x_3\sigma_{12},\;
(\sigma_{ik} = \sigma_i \sigma_k).$
The reduced form of the EMA equation is
$$
y^3 + py + q = 0,
\en(C2)
$$
where $y = \sigma_e + a_1/3, \; p=-a_1^2/3 + a_2,\; q=2(a_1/3)^3-a_1 a_2/3+a_3.$
Three solutions of the equation (C2) are
$$
y_1=A+B, \quad y_{2,3}= -(A+B)/2 \pm i (A-B)/2 \sqrt{3},
$$
$$
A=\left(-\fr{q}{2}+\sqrt{Q}\right)^{1/3}, \quad
B=\left(-\fr{q}{2}-\sqrt{Q}\right)^{1/3}, \quad
Q=(\fr{p}{3})^3 + (\fr{q}{2})^2.
\en(C3)
$$
In order to find the physical solution let us consider firstly the homogeneous
limit $\sigma_1 = \sigma_2 = \sigma_3 = \sigma.$ Then, using the normalization
condition for concentrations $\sum_{i=1}^3 x_i = 1$ and the degenerated
values of $s_i$ ($s_1 =3\sigma,\; s_2 = 3\sigma^2, s_3 = \sigma^3$) one can
see that
$$
a_1= \sigma,\; a_2 = -\sigma^2,\; p=-\fr{4}{3} \sigma^2, \; q =-2(\fr{2}{3})^3 \sigma^3
\quad Q = 0,\; A=B=\fr{2}{3} \sigma,
\en(C4)
$$
and consequently
$$
y_1 = \fr{4}{3}\sigma, \quad y_{2,3} = -\fr{2}{3}\sigma,\quad
\sigma_{e1} = \sigma, \quad \sigma_{e2,3} = -\sigma.
\en(C5)
$$
Thus, the solution $y_1$ corresponds to the physical one.
The expression (C3) has a complicated form in general case.
Let us consider the physical solution at some special values of parameters,
when it takes more simple, tractable form.

1) Equal phase concentrations, $x_i = x = 1/3.$

In this case all coefficients can be expressed only through $s_i.$
Then one obtains  for $y_1$
$$
a=s_1/3, \; b=-s_2/3, \; p=(-s_1^2 +9s_2)/27, \;
q=(2s_1^3-27s_1 s_2 -27^2 s_3)/(27)^2,
$$
$$
Q= (729 s_3^2 - 4s_3 s_1^3 - 54 s_3 s_2 s_1 - s_1^2 s_2^2/3 -4s_2^3)/4 {27}^2.
\en(C7)
$$
Even in this case it is not so simple to check
that $\sigma_e$ satisfies the inversion relation.
The equality of the numerical coefficients before $a_1$ and $a_2$ is a consequence
of the duality relation. The simple result for $\sigma_e,$ coinciding with the
exact values (16-18), can be obtained only in the corresponding FP
$\sigma_i\sigma_j = |\epsilon_{ijk}|\sigma_k^2, \; x_i=x_j, \;(i,j,k = 1,2,3),$
where $\epsilon_{ijk}$ is the fully antisymmetric tensor and there is no summation
on $(i,j,k).$
2) To see this let us consider the FP with arbitrary equal concentrations of two
first phases:
$\sigma_1\sigma_2 = \sigma_3^2,\; x_1=x_2=x.$
In this case
$$s_1 = \sigma_1 + \sigma_2 + \sqrt{\sigma_1\sigma_2},\quad
s_2= \sqrt{\sigma_1\sigma_2}s_1, \quad s_3 = (\sigma_1\sigma_2)^{3/2},
$$
$$
\bar s_1 = \bar \sigma = x(\sigma_1 + \sigma_2) + (1-2x)\sqrt{\sigma_1\sigma_2},\quad
\ti s_2 = \sqrt{\sigma_1\sigma_2} \bar \sigma,
$$
$$
a_1 = s_1 - 2\bar \sigma\;
a_2 = \sqrt{\sigma_1\sigma_2} (-s_1 + 2\bar \sigma) = - a_1 \sqrt{\sigma_1\sigma_2}.
$$
Thus one see that the coefficients of equation (C1) satisfy the fixed point
relations (42) for arbitrary $x.$  It means that eq.(C1) has the solution
$f=1, \; \sigma_e = (s_3)^{1/3} = \sqrt{\sigma_1\sigma_2} = \sigma_3.$
The similar behaviour takes place for other fixed points.
\bs

\cl{\bf Acknowledgments}

\bs
The author is thankful to V.Marikhin for useful information about heterophase systems
and to Prof. F.Kusmartsev for useful discussions and a warm hospitality at the
Loughborough University, UK.
This work was supported by RFBR grants \# 2044.2003.2 and \# 02-02-16403.

\end{document}